\documentclass{iaus}
\usepackage{graphicx}
\usepackage{natbib}

\def\noprintlabel{}
\def\biblab #1{\ifx\noprintlabel\undefined{\bf [#1]}\fi}
\def\note #1]{{\bf #1]}}
\def\rt{r_{\rm t}}
\def\rc{r_{\rm c}}
\def\psit{\psi_{\rm t}}
\def\cm{\,{\rm cm}}
\def\s{\,{\rm s}}
\def\dd{{\rm d}}
\def\CK{{\cal K}}
\def\Rsun{R_\odot}
\def\fig{.}
\def\etal{{\it et~al.\/}}

\title[Rotation of the Solar Convection Zone] 
{Rotation of the Solar Convection Zone from Helioseismology}

\author[J. Christensen-Dalsgaard]   
{J{\o}rgen Christensen-Dalsgaard$^1$}%

\affiliation{$^1$
Institut for Fysik og Astronomi,
Aarhus Universitet, DK-8000 Aarhus C, Denmark
\break email: jcd@phys.au.dk}

\pubyear{2007}
\volume{239}  
\pagerange{119--126}
\date{?? and in revised form ??}
\jname{Convection in Astrophysics}
\editors{F. Kupka, I.W. Roxburgh \& K.L. Chan, eds}
\begin{document}

\maketitle

\begin{abstract}
Helioseismology has provided very detailed inferences about rotation
of the solar interior. Within the convection zone the 
rotation rate roughly shares the latitudinal variation seen in the
surface differential rotation.
The transition to the nearly uniformly
rotating radiative interior takes place in a narrow tachocline, which is
likely important to the operation of the solar magnetic cycle.
The convection-zone rotation displays zonal flows,
regions of slightly more rapid and slow rotation, extending over much
of the depth of the convection zone and converging towards the equator
as the solar cycle progresses. In addition, there is some evidence for
a quasi-periodic variation in rotation, with a period of around 1.3 yr,
at the equator near the bottom of the convection zone.
\end{abstract}

\firstsection 
\section{Introduction}

Convection and rotation are intimately linked in the solar convection zone.
The convective timescales in the deeper parts of the convection zone are
similar to the solar rotation period, giving rise to 
a strong influence
of rotation on the convection dynamics.
Also, transport of angular momentum by convection is likely
important for the solar surface differential rotation
and the interplay between convection and rotation presumably also
causes other dynamical phenomena in the convection zone, such as
the meridional circulation.
Finally, interaction between convection, rotation and other flows are 
assumed to lead to the generation of the solar large-scale magnetic fields
and their cyclic variation.

Before the advent of helioseismology, observation of solar rotation was limited
to the motion of features across the solar disk, or measurements of the
Doppler velocity of the solar surface plasma.
This showed the variation in rotation with latitude, from a rotation period
around 25 days at the equator to around 35 days near the pole,
although the motion at high latitudes was difficult to follow.
Also, differences in rotation rate between the surface plasma and
magnetic features presumed to be
anchored at different depths gave some indication of an increase of
angular velocity with depth in the near-surface region
\citep[e.g.,][]{Wilcox1970, Foukal1972}.

Theoretical modelling through hydrodynamical simulations of the convection
zone \citep{Glatzm1985, Gilman1986}
was able to reproduce the surface variation and tended to show an angular
velocity depending only on the distance from the rotation axis;
such `rotation on cylinders' was expected from simple hydrodynamical
arguments, leading to the Taylor-Proudman theorem \citep{Pedlos1987}.
Modelling of rotation of the deep solar interior was extremely uncertain.
From observations of other stars it is assumed that the Sun was rotating
substantially more rapidly in its early life, with a loss of angular momentum
through the magnetized solar wind coupled to the convection zone;
however, it was unclear to what extent transport of angular momentum in the
radiative interior would lead to a spin-down of the solar core.



The observational situation has been changed dramatically by helioseismology.
Already early analyses by \cite{Deubne1979} of high-degree five-minute
oscillations showed the effect of the advection of the modes by rotation,
providing an independent indication of an increase of the velocity with depth.
\cite{Gough1983} provided a more solid theoretical basis for the extraction
of such subsurface velocity signals or sound-speed fluctuations from
helioseismic data.
With observations over the last decade from the GONG network \citep{Harvey1996}
and the MDI instrument on the SOHO spacecraft \citep{Scherr1995}
we now have a detailed picture of rotation in the solar interior and its 
variation over a full 11-year cycle.
Interestingly, this does not conform with the early models 
of rotation on cylinders in the convection zone;
as discussed by Brun {\etal}\ (these proceedings) models of 
the convection zone are still not fully able to account for the 
inferred rotation profile.
A striking feature is the sharply localized change from the latitudinally
differential rotation in the convection zone to the nearly solid-body 
rotation in the radiative interior.
This takes place in a narrow region, the so-called {\it tachocline\/}
\citep{Spiege1992} near the base of the convection zone,
of likely substantial importance to the generation of the solar magnetic
cycle.

Here I give a brief overview of the helioseismic analyses that have led to
these inferences of solar internal rotation and discuss some of the results.
A more extensive review of solar rotation and helioseismic investigations
of it was given by \cite{Thomps2003}, while
\cite{Christ2007} discussed the observed
properties of the tachocline region in more detail.
A recent detailed review of observation and modelling of the dynamics of the
solar convection zone was provided by \cite{Miesch2005}.


\section{Helioseismic inferences of rotation}\label{sec:helioseis}

%
As a background for interpreting the helioseismic results a short introduction
to helioseismology is probably useful.
Extensive treatments of stellar oscillations have been provided by
\cite{Unno1989} and \cite{Gough1993},
while \cite{Christ2002} reviewed the techniques and results of helioseismology.

\subsection{Properties of solar oscillations}\label{sec:oscprop}

Solar oscillations are believed to be excited stochastically by 
near-surface convection. 
They are observed at periods between roughly fifteen and three minutes, with
largest amplitudes near periods of five minutes.
The amplitudes per mode are minute:
at most around $20 \cm \s^{-1}$ in radial velocity
and below a few parts per million in intensity.
Even so, it has been possible to determine the mode frequencies with extremely
high accuracy, thus forming the basis for the helioseismic analyses.


A mode of solar oscillations depends on co-latitude $\theta$ and longitude
$\phi$ as a spherical harmonic, characterized by the {\it degree\/} $l$ 
and the {\it azimuthal order\/} $m$, with $|m| \le l$.
In addition, the mode is characterized by its {\it radial order\/} $n$
which provides a measure of the number of nodes in the radial direction.
The angular frequency $\omega_{nlm}$ in general depends 
on all three wave numbers. 

The observed modes of solar oscillations are acoustic (or p) modes or, at 
relatively high degree, surface gravity (or f) modes.
The p modes are trapped between the photosphere and an inner turning 
point at a distance $\rt$ from the centre, determined by
\begin{equation}
{c (\rt) \over \rt} = {\omega \over \sqrt{l(l+1)}} \; ,
\label{eq:rt}
\end{equation}
where $c$ is the adiabatic sound speed.
Thus high-degree modes are trapped near the surface while low-degree modes
penetrate to the solar core.
For the f modes the displacement amplitude decreases exponentially with
depth below the surface with an e-folding distance of approximately
$\Rsun/l$, $\Rsun$ being the solar radius.
The extent of the eigenfunctions in latitude is determined by the properties
of the spherical harmonics; asymptotically, a mode is confined between
latitudes of $\pm \psit$, 
given by
\begin{equation}
\cos \psit = {m \over l+1/2} \; ;
\label{eq:latt}
\end{equation}
thus modes with $m \simeq l$ are confined near the equator, whereas modes
with $m/l \ll 1$ extend over all latitudes.

\begin{figure}
 \centerline{
 \scalebox{0.555555}{
 \includegraphics{\fig/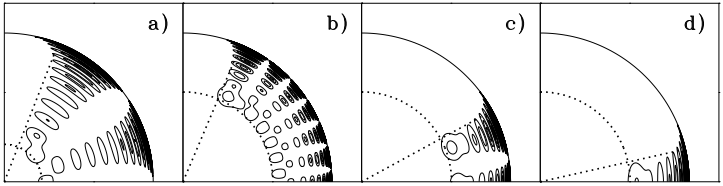}
 }}
  \caption{
  Contour plots of rotational kernels $K_{nlm}$ in a solar quadrant.
  The modes all have cyclic frequencies $\omega/2 \pi$
  near 2~mHz; the following pairs of
  $(l, m)$ are included: a) $(5,2)$; b) $(20,8)$; c) $(20,17)$; and
  d) $(20,20)$.
  The dotted circles mark the locations of the lower radial turning
  point $\rt$ (cf.\ eq. \ref{eq:rt})
  and the dotted lines show the latitudinal turning
  points $\psit$ (cf.\ eq. \ref{eq:latt}).
  \label{fig:kernels}
    }
\end{figure}
%


The Sun is rotating so slowly that the centrifugal force or other higher-order
effects of rotation can be neglected. 
For the observed modes the dominant effect of rotation on the frequencies 
is simply the advection of the modes by rotation:
the dependence of an oscillation on longitude
and time $t$ can be written as $\cos(\omega_{nlm} t - m \phi)$, i.e.,
for $m \neq 0$ behaving as a wave running in the longitude direction;
thus the frequencies
of modes travelling in the direction of rotation are increased and frequencies
of modes 
travelling in the direction opposite to rotation are decreased.
It is plausible that the resulting frequencies are given by
$\omega_{nlm} \simeq \omega_{nl0} + m \langle \Omega \rangle$,
where $\langle \Omega \rangle$ is a suitable average, determined by the 
properties of the mode, of the angular velocity $\Omega(r, \theta)$;
as indicated, 
$\Omega$ is in general a function of the distance $r$ to the centre and
$\theta$.
A more careful analysis shows that the {\it rotational splitting\/} can
be written as
\begin{equation}
\delta \omega_{nlm} \equiv \omega_{nlm} - \omega_{nl0}
= m \int_0^{\Rsun} \int_0^\pi K_{nlm}(r, \theta) \Omega(r, \theta) 
r \dd r \dd \theta \; ,
\label{eq:rotsplit}
\end{equation}
where the {\it kernels\/} $K_{nlm}$ are determined by the structure of the Sun
and the properties of the eigenfunctions.
It is important to note that the kernels are symmetrical around the equator;
thus the rotational splitting is only sensitive to the similarly symmetric
component of $\Omega(r, \theta)$, and analysis of rotational splittings of
global modes provides no information about the antisymmetric component of
$\Omega$.
Some examples of kernels are illustrated in Figure~\ref{fig:kernels};
it is evident that they largely follow the behaviour expected from the 
turning points in $r$ and $\theta$.

\subsection{Inverse analysis}

%
By observing a large selection of modes of such varying extent it is
possible to make combinations of the rotational splittings that
provide localized measures of $\Omega(r, \theta)$.
Techniques for such {\it inverse analyses\/} were discussed by
\cite{Schou1998}.
In many cases the inference of $\Omega(r_0, \theta_0)$ at some
point $(r_0, \theta_0)$ is obtained as a linear combination of the 
data which, for example, could be taken on the form
$m^{-1} \delta \omega_{nlm}$:
\begin{equation}
\Omega(r_0, \theta_0) = 
\sum_{nlm} c_{nlm}(r_0, \theta_0) m^{-1} \delta \omega_{nlm} 
= \int_0^{\Rsun} \int_0^\pi \CK(r_0, \theta_0, r, \theta) \Omega(r, \theta) 
r \dd r \dd \theta \; ,
\label{eq:rotinv}
\end{equation}
by using equation (\ref{eq:rotsplit}).
Here the {\it averaging kernel\/} is given by
\begin{equation}
\CK(r_0, \theta_0, r, \theta) = 
\sum_{nlm} c_{nlm}(r_0, \theta_0) K_{nlm}(r, \theta) \; ;
\label{eq:avker}
\end{equation}
it is typically defined such as to have unit integral over $(r, \theta)$.
Thus equation (\ref{eq:rotinv}) defines the inferred angular velocity
as an average which ideally is localized near $(r_0, \theta_0)$;
the degree of localization is determined by the extent of the averaging kernel.
Also, the error in $\Omega(r_0, \theta_0)$ can obviously be 
determined from equation (\ref{eq:rotinv}) if the error properties of the
data are known.
How the {\it inversion coefficients\/} $c_{nlm}(r_0, \theta_0)$ are obtained
depends on the properties of the specific inversion method;
in some techniques they are found so as explicitly to localize the
averaging kernel, whereas other techniques attempt to fit the inferred
angular velocity to the observations in a least-squares sense
(see \cite{Schou1998} for details).

\begin{figure}
 \centerline{
 \scalebox{0.35}{
 \includegraphics{\fig/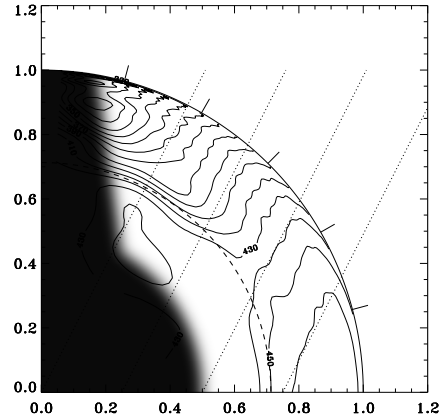}
 }}
  \caption{
  Inferred rotation rate $\Omega/2 \pi$ in a quadrant of the Sun,
  obtained by means of inversion of 144 days of MDI data.
  The equator is at the horizontal axis and the pole is at the vertical
  axis, both axes being labelled by fractional radius.
  Some contours are labelled in nHz, and, for clarity, selected
  contours are shown as bold. The dashed circle is at the base of
  the convection zone and the tick marks at the edge of the outer
  circle are at latitudes $15^\circ$, $30^\circ$, $45^\circ$, $60^\circ$, and
  $75^\circ$.
  The shaded area indicates the region in the Sun where no reliable inference
  can be made with the present data.
  The slanted dotted lines are at an angle of $27^\circ$ with the
  rotation axis.
  \citep[Adapted from][]{Schou1998}.
  \label{fig:rotation}
    }
\end{figure}

\section{The solar internal rotation}

%
The inversion techniques discussed above have been extensively applied to
observed solar oscillation frequencies.
Already early results \citep{Duvall1984} clearly demonstrated that
the radiative interior rotated at roughly the surface rate, with little or 
no indication of a rapidly rotating core.
With the determination of rotational splittings for different values of $m$
it became possible to obtain information about the variation of the
latitudinally differential rotation with depth
\citep{Brown1987, Christ1988, Brown1989, Dziemb1989},
demonstrating the transition between differential rotation in the convection 
zone and near-constant rotation in the radiative interior.
Extensive analyses by \cite{Schou1998} have showed a considerable degree
of consistency between results of different inversion methods, although
some sensitivity remains to the techniques used in the basic analysis
to determine the rotational splittings from the observations \citep{Schou2002},
particularly at high latitudes.
Using also data on low-degree modes from unresolved observations the 
absence of a rapidly rotating core has been confirmed, although the
inferences are still uncertain 
in the inner around 20 \% of the solar radius \citep[e.g.,][]{Chapli1999}.
This confirms that an efficient mechanism must have linked the radiative
interior to the loss of angular momentum from the convection zone;
the nature of this mechanism is still debated, however.

\subsection{The bulk of the convection zone}

\begin{figure}
 \centerline{
 \scalebox{0.6}{
 \includegraphics{\fig/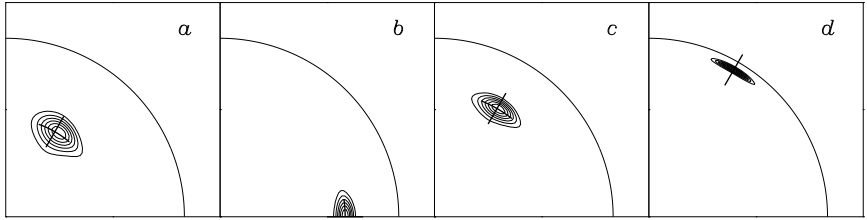}
 }}
  \caption{
  Averaging kernels for the inversion shown
  in Figure \ref{fig:rotation},
  targeted at the following radii and latitudes in the Sun:
  ($0.55\Rsun$, $60^\circ$),
  ($0.7\Rsun$, $0^\circ$), ($0.7\Rsun$, $60^\circ$), 
  and ($0.95\Rsun$, $60^\circ$).
  The corresponding locations are indicated with crosses.
  \citep[Adapted from][]{Schou1998}.
  \label{fig:avker}
    }
\end{figure}

%
Here I concentrate on rotation within and just below the convection zone.
Figure \ref{fig:rotation} shows an inferred rotation profile in the
outer parts of the Sun.
Some averaging kernels for this inversion are illustrated 
in Figure~\ref{fig:avker};
evidently the combinations of the observed splittings are able to
determine the angular velocity with fairly high resolution, particularly
in the radial direction near the surface.
In the bulk of the convection zone the latitudinal variation clearly
roughly follows the variation at the surface, the helioseismically
determined values close to the surface being essentially consistent with direct 
Doppler measurements of the surface rotation \citep[e.g.,][]{Beck2000}.
It is obvious that the rotation is not constant on cylinders, 
as simple models predicted.
As noted by \cite{Gilman2003} it rather appears that in much of
the convection zone rotation is constant
on lines inclined at $27^\circ$ to the rotation axis;
the physical reason for this property is not understood.

\begin{figure}
 \centerline{
 \scalebox{0.25}{
 \includegraphics[angle=90]{\fig/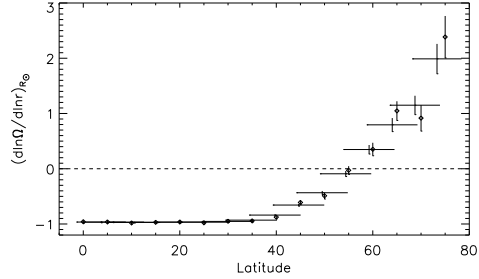}
 }}
  \caption{
Logarithmic gradient of the angular velocity obtained from linear fits
in the outer two per cent of the solar radius
to the results of inversion of f-mode splittings.
  \citep[Adapted from][]{Corbar2002}.
  \label{fig:surshear}
    }
\end{figure}

Very near the surface the contours clearly indicate that the rotation
rate decreases outwards, at low and intermediate latitude.
This corresponds to the behaviour inferred from surface markers
presumed to be anchored at different depths and is reminiscent of
the early results of \cite{Deubne1979}.
In a more careful analysis based on f modes \cite{Corbar2002}
determined the gradient of the angular velocity in this near-surface region,
as shown in Figure \ref{fig:surshear}.
Interestingly, $\dd \ln \Omega / \dd \ln r \simeq -1$ at latitudes
below around $40^\circ$;
this is not consistent with the, perhaps natural, assumption of constant
angular momentum maintained in this region by convective motions,
which would instead correspond to $\dd \ln \Omega / \dd \ln r = -2$.

\begin{figure}
 \centerline{
 \scalebox{0.5}{
 \includegraphics{\fig/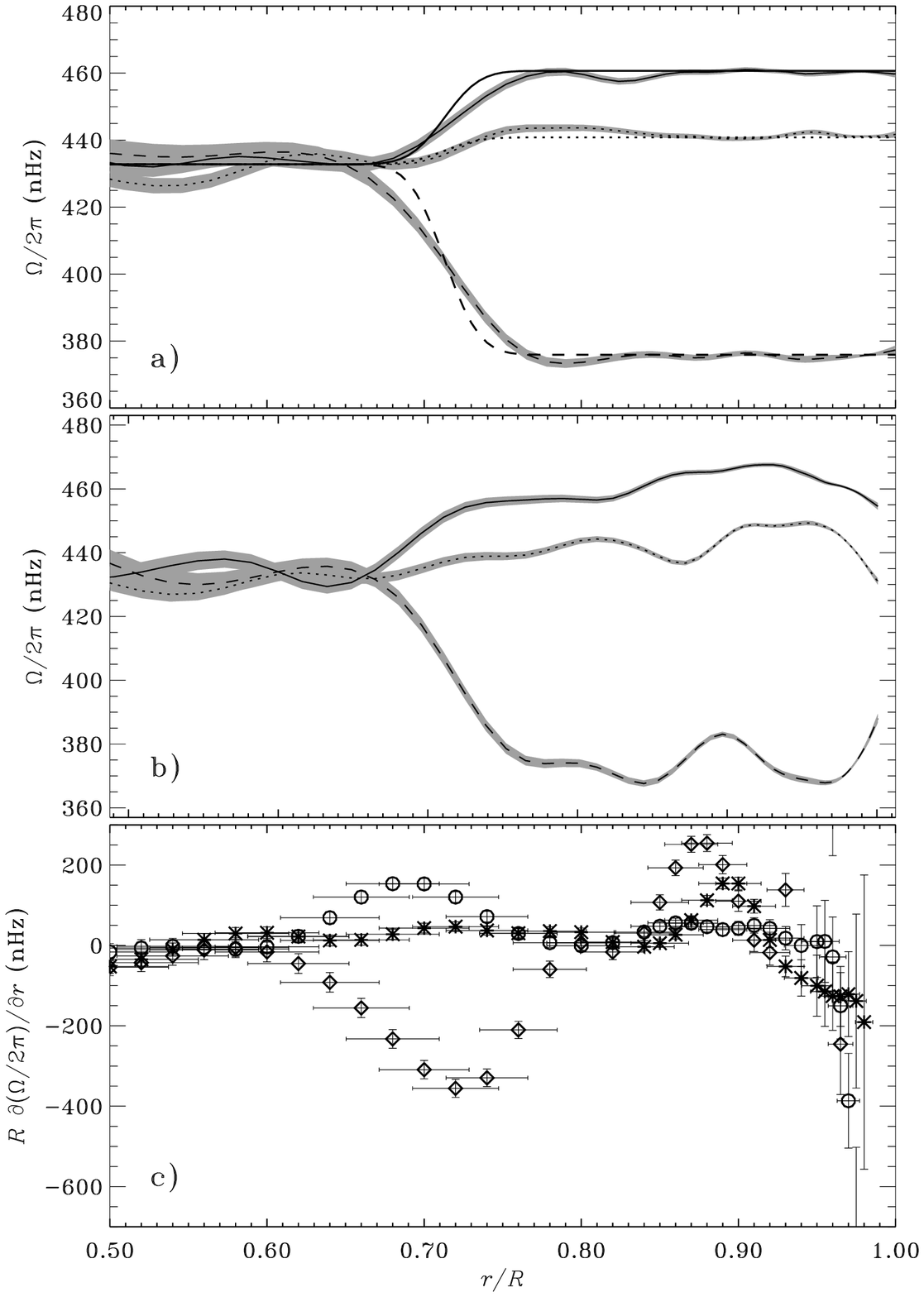}
 }}
\vskip -1cm
  \caption{
(a) Inversion of synthetic dataset for artificial rotation profile,
based on artificial data with properties corresponding to an observed dataset.
The thick curves show the underlying input rotation profile, in the form
of radial cuts at latitudes of 60$^\circ$ (dashed curve),
30$^\circ$ (dotted curve), and 0$^\circ$ (equator; solid curve).
The thin curves show a typical inversion solution, with
the $1 \sigma$ range, determined from the errors in the observed 
mode set, indicated by the shaded areas.
(b) Corresponding inversion result for the solar data.
(c) Results of inversion for the radial gradient of the solar rotation 
rate at latitude $0^\circ$ (circles),
$30^\circ$ (asterisks) and $60^\circ$ (diamonds);
here the vertical error bars indicate $1 \sigma$ errors,
whereas the horizontal bars indicate the resolution of the inversion
as determined by the extent in radius of the averaging kernels.
\citep[Adapted from][]{Charbo1999}.
  \label{fig:tachocline}
    }
\end{figure}

\subsection{Properties of the tachocline}

%
As already mentioned, the properties of the tachocline are of very 
considerable interest.
Particularly important are the location and thickness of the transition,
first addressed in detail by \cite{Kosovi1996}.
Determination of these properties
must obviously take into account the resolution of the inversion,
as characterized by the averaging kernels (cf.\ Figure~\ref{fig:avker}).
The effect on the inversion is illustrated in Figure~\ref{fig:tachocline}a,
on the basis of inversion of artificial data 
computed for the rotation profile shown by the heavy curves.
The corresponding inversion results, indicated by thin curves and shaded
areas, clearly show a transition that is more gradual than for the `true'
underlying rotation rate, as a result of the smearing by the averaging kernel.
The result of the corresponding inversion of solar data
\citep{Charbo1999}, shown in Figure~\ref{fig:tachocline}b,
in fact indicates a width that roughly corresponds to the width assumed
for the artificial data.
Further information about the transition can be obtained,
as discussed by Charbonneau {\etal},
by arranging the linear combination of the data in equation (\ref{eq:rotinv})
such that the corresponding combination $\CK$ of the kernels provides
an average of the radial gradient of $\Omega$.
The result (Figure~\ref{fig:tachocline}c)
clearly shows the localized gradient in the tachocline,
furthermore hinting that the transition takes place closer to the surface at
latitude $60^\circ$ than at the equator.

From tests on artificial data, and from parametrized fits,
it is to some extent possible to correct for the finite resolution
\citep[e.g.,][]{Charbo1999, Corbar1999}, and hence obtain a measure
of the actual width.
For example, \cite{Basu2003} determined the location $\rc$ of the midpoint
of the transition and its width $w$%
\footnote{defined such that 84 \% of the transition takes place
between $\rc - w$ and $\rc + w$}
as $\rc = 0.692 \Rsun$, $w = 0.033 \Rsun$ at the equator
and $\rc = 0.710 \Rsun$, $w = 0.076 \Rsun$ at latitude $60^\circ$.
Thus they confirmed the prolate nature of the transition and furthermore
found that it is broader at high latitude than at the equator.
For comparison, the radius at the base of the convection zone,
assuming spherical symmetry, has been determined from helioseismology
as $0.713 \Rsun$ \citep[e.g.,][]{Christ1991};
thus, under this assumption, most of the transition takes place 
below the convection zone at the equator, whereas at higher
latitude there is considerable overlap between the tachocline and
the convection zone.

%


\section{Variations with time}

Doppler observations of the solar surface rotation have shown variations
with time in the so-called {\it torsional oscillations\/}
\citep{Howard1980}, bands of slightly faster and slower rotation that 
converge towards the equator as the solar cycle progresses.
The availability of detailed helioseismic observations extending over
a sunspot cycle now allows investigations of variations
in solar structure, rotation and other types of dynamical phenomena
beneath the solar surface.
Particularly detailed investigations have been made of changes in rotation;
the principle is to analyse the data in segments of typically 2 -- 3 months,
and consider the residual from the time-averaged rotation rate, as
a function of $r$ and $\theta$.
A recent overview of the results was provided by \cite{Howe2006r}.

\subsection{Tachocline oscillations?}

\begin{figure}
 \centerline{
 \scalebox{0.5}{\includegraphics{\fig/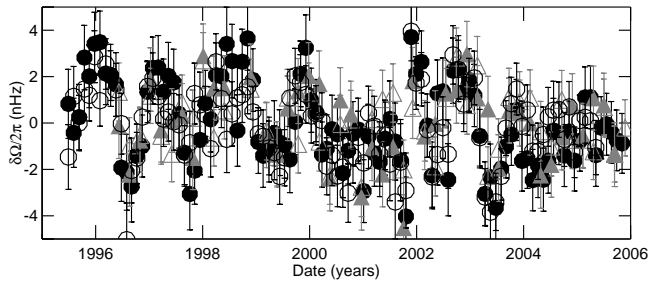}}
}
  \caption{
Residuals from average rotation rate at $r = 0.72 \Rsun$ at the equator.
Triangles and circles are based on MDI and GONG data, respectively, and
open and closed symbols correspond to two different inversion techniques.
  \citep[Adapted from][]{Howe2007}.
  \label{fig:tachoosc}
    }
\end{figure}

Perhaps the most surprising result of such analyses was the detection of
what appeared to be an oscillation with a period of 1.3~yr in the equatorial
rotation rate near the base of the convection zone \citep{Howe2000a}.
As shown in Figure~\ref{fig:tachoosc}, this oscillation was present,
with largely consistent behaviour, in analyses of two different datasets and
using two different inversion methods, between late 1995 and 2000, while
the variation has been more erratic since then.
The early period also appeared to show an oscillation with a similar period,
but the opposite phase, at somewhat greater depth at the equator.
Given the subtle nature of the signal, it is important to check whether
it could be an artefact of, for example, 
variations in the selection of modes included in the analysis;
this appears not to be the case \citep{Toomre2003}.
However, \cite{Basu2001}, while apparently detecting a similar variation,
questioned its statistical significance.
It is evident that the erratic nature of the variation may cast doubt
on its physical reality;
on the other hand, it is not implausible that such behaviour, of possibly
magnetic origin, could be linked to specific phases in the solar cycle.
Thus continued observations during the rising phase of the next cycle will
be of great interest.

\subsection{Zonal flows}

Helioseismic investigations have showed that, far from being a
superficial phenomenon, the torsional oscillations seen on the solar
surface \citep{Howard1980} extend quite deeply.
From inversion of f modes
\cite{Kosovi1997} found that the regions of slightly faster and
slower rotation extended a few per cent of the solar radius
beneath the solar surface.
\cite{Schou1999} considered f-mode data covering more than two years
and found that these subsurface zonal flows shared the propagation 
towards the equator seen on the surface, closely linked to the `butterfly
diagram' of the sunspots.
More extensive data, on both p and f modes, were considered by
\cite{Antia2000} and \cite{Howe2000b}, who confirmed the
equator-ward propagation of the zonal flows at low and intermediate latitude,
the flow extending through up to one third of the convection-zone depth.
At high latitude, on the other hand, flows of somewhat higher amplitude
propagated towards the pole.
Even deeper penetration, particularly at high latitude,
was inferred by \cite{Voront2002}, using a non-linear inversion technique.
They also found that the propagation could be fitted as having an 
11-yr period and determined the amplitude and phase of this variation.

\begin{figure}
 \centerline{
 \scalebox{0.38}{\includegraphics{\fig/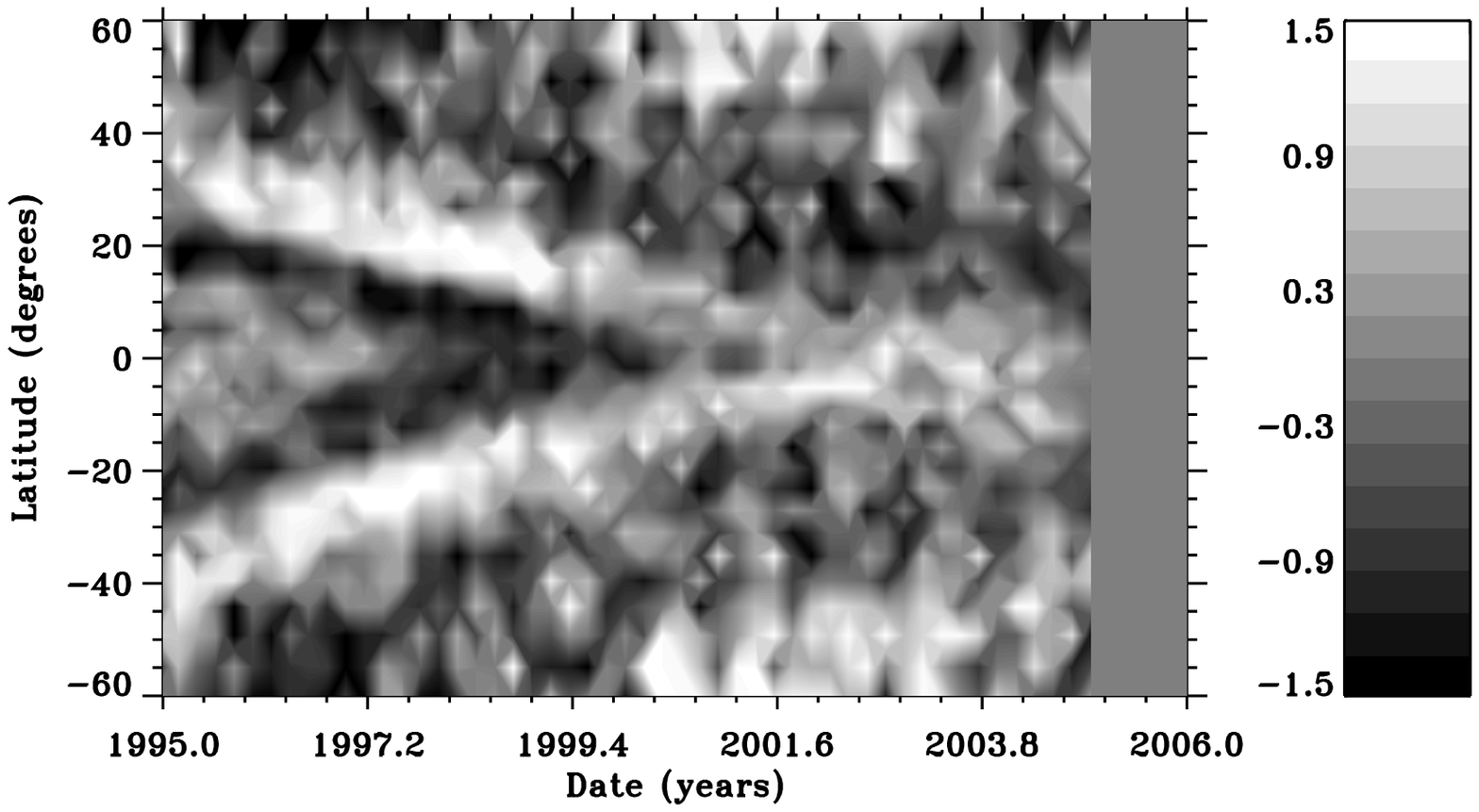}}
}
\centerline{
 \scalebox{0.38}{\includegraphics{\fig/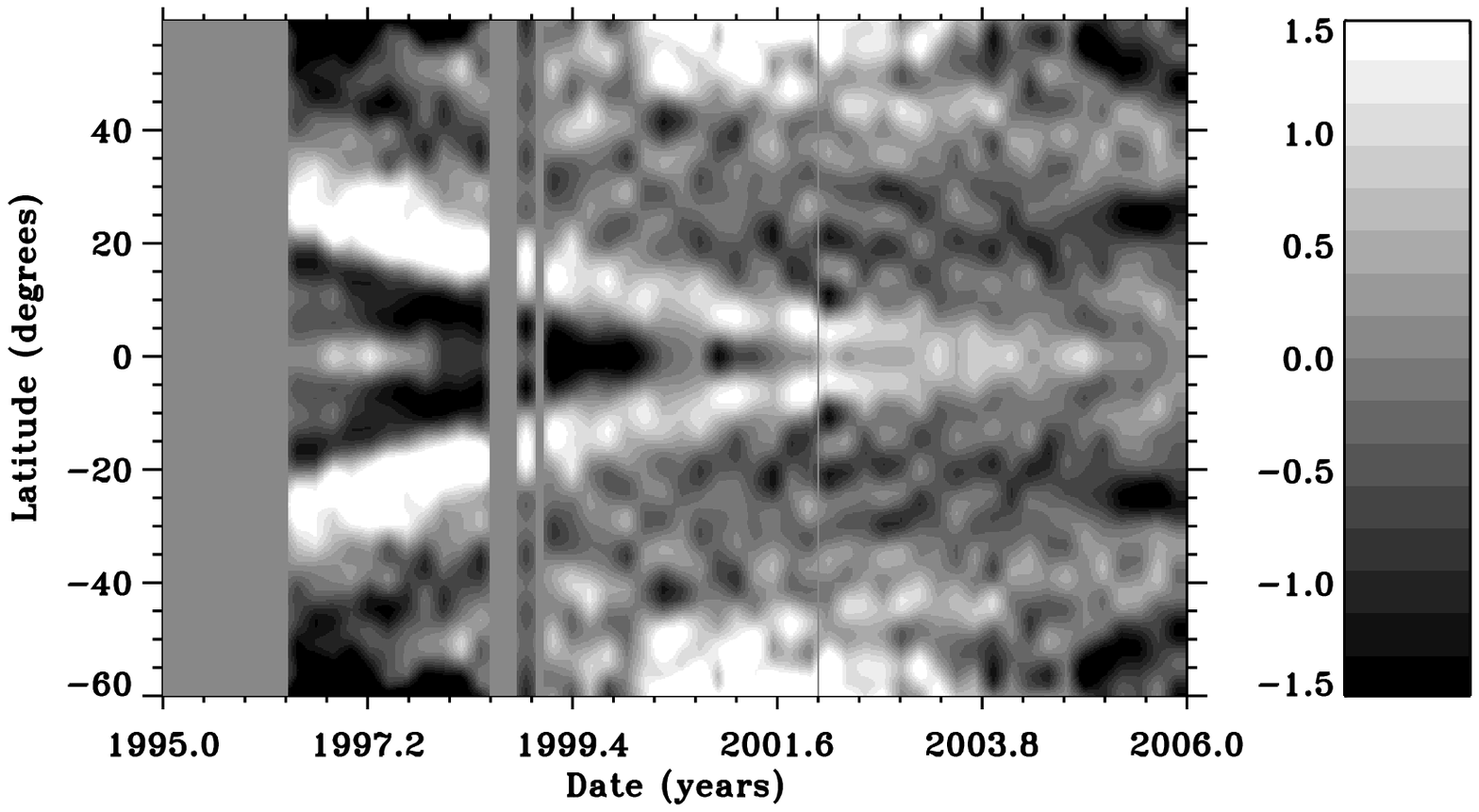}}
 \scalebox{0.38}{\includegraphics{\fig/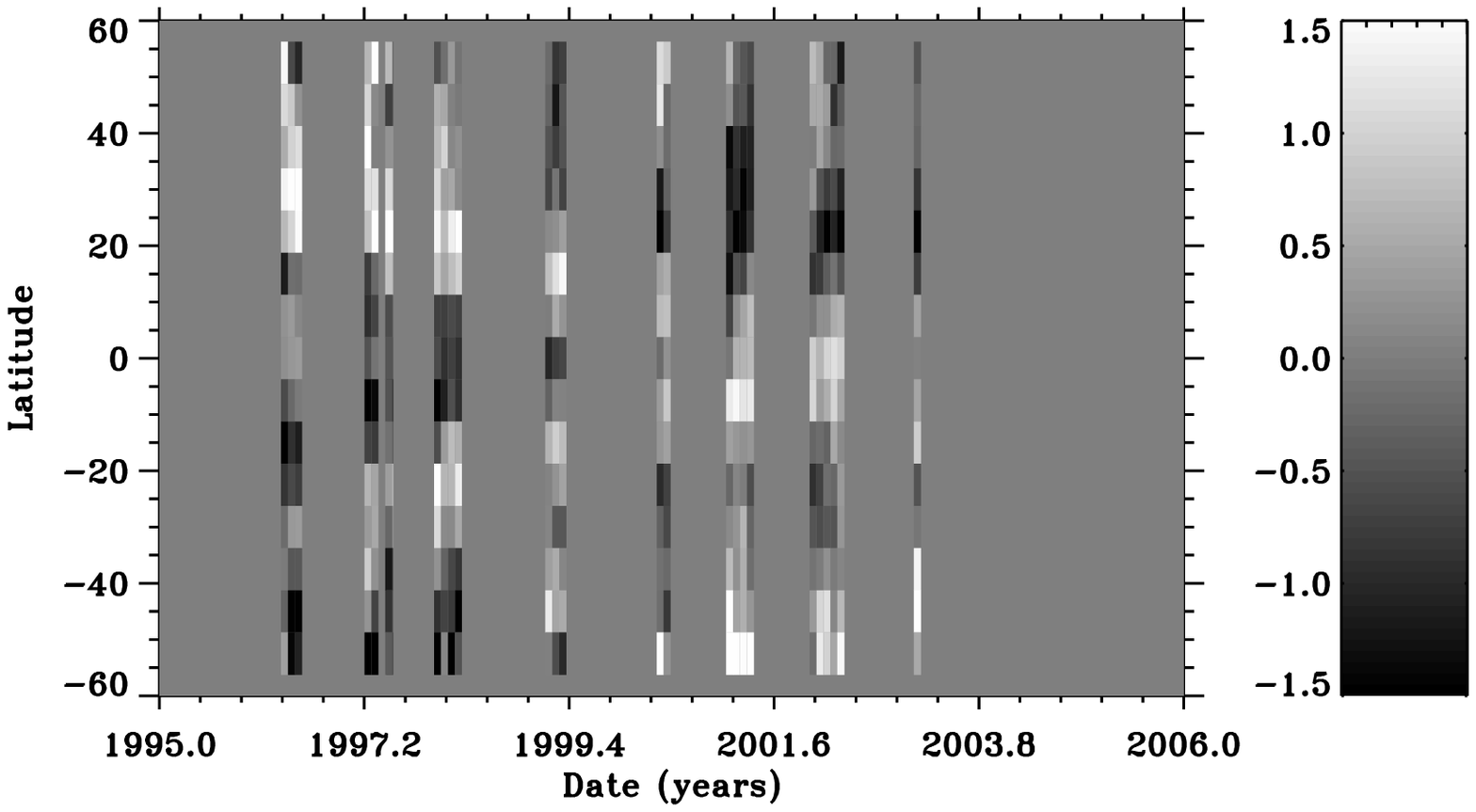}}
}
  \caption{
Zonal flow patterns, obtained as residuals from time-averaged rotation rates.
The top panel shows the `torsional oscillations' obtained from Mount Wilson
surface Doppler observations.
The two lower panels are from inverse analyses targeted at $r = 0.99 \Rsun$;
in the left panel global inversion of MDI data was used, whereas
the right panel is based on a local helioseismic analysis with the
ring-diagram technique (see Section~\ref{sec:local}),
during time periods where high-resolution MDI data were available.
  \citep[Adapted from][]{Howe2006a}.
  \label{fig:zonallat}
    }
\end{figure}

\begin{figure}
 \centerline{
 \scalebox{0.6}{\includegraphics{\fig/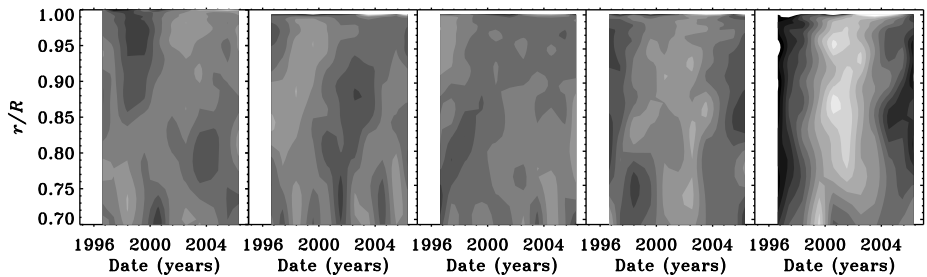}}
}
  \caption{
Rotation residuals as a function of time and fractional distance 
from the centre, from inversions of MDI data, at latitudes
(left to right)
$0^\circ$, $15^\circ$, $30^\circ$, $45^\circ$ and $60^\circ$.
The data have been averaged over periods of 1~yr.
Darker regions correspond to rotation slower than average, and 
lighter regions to faster rotation.
  \citep[Adapted from][]{Howe2006b}
  \label{fig:zonalrad}
    }
\end{figure}

As an example of these results, Figure~\ref{fig:zonallat} compares flows 
from Doppler observations at the solar surface with helioseismically
inferred flows at a depth of $0.01 \Rsun$.
The two patterns are evidently very similar.
The variation with depth is illustrated in Figure~\ref{fig:zonalrad};
this confirms that the flows can be followed through most of the 
convection zone.
Also, strikingly, the flow pattern appears to propagate towards the solar
surface as time progresses; this is difficult to reconcile with a model
proposed by \cite{Spruit2003} according to which the flow originates
from thermal effects at the solar surface.

%


The origin of these zonal flows has been discussed in the context
of mean-field dynamo models 
\citep[e.g.,][]{Covas2000, Rempel2006a, Rempel2006b}.
Rempel found that the pole-ward propagating branch could be explained
by the feed-back of the Lorentz force on differential rotation in such models.
He also showed that a similar mechanical model for
the equator-ward propagating low-latitude branch would be strongly 
affected by the Taylor-Proudman theorem, leading to a variation
with depth and latitude that is inconsistent with observations;
thus he concluded that thermal effects must be involved.

\section{Local helioseismology}\label{sec:local}

I have so far considered {\it global\/} helioseismology, based on frequencies
of global modes of solar oscillation.
This is restricted to consider only rotationally symmetric aspects of
the solar interior structure and dynamics, such as rotation;
furthermore, as noted in Section~\ref{sec:oscprop}, the analysis is
sensitive only to aspects that are symmetrical around the equator.
These restrictions can be avoided through the use of 
{\it local helioseismology\/} where the data are analysed instead in terms
of wave propagation in a smaller part of the solar surface.
An extensive review of the techniques and results of such analyses
was given by \cite{Gizon2005}.

Detailed investigations of sub-surface flows have been made 
with the {\it ring-diagram technique\/}, following the
early analysis of \cite{Gough1983}.
The oscillations are analysed in smaller patches on the solar surface,
to produce local power spectra in terms of the horizontal wave vector
${\bf k}_{\rm h}$ \citep{Hill1988}.
By considering a large number of such patches distributed over the 
solar disk one can build up a picture of the subsurface structure and
flows as a function of position on the Sun.
A detailed investigation of this nature was carried out by
\cite{Haber2002}; they determined the longitudinally averaged 
rotational flow, also identifying the zonal flows. 
The results were similar to those of the global analyses,
but with clear differences between the northern and southern hemispheres.
This is also illustrated in Figure~\ref{fig:zonallat}, where the lower
right panel shows results obtained with a ring-diagram analysis.
Similar results were obtained by \cite{Zhao2004} using the
{\it time-distance technique\/}, where a correlation analysis is used
to infer the travel time of waves between different points on the
solar surface.

Local helioseismology has also revealed other kinds of large-scale 
flows beneath the solar surface.
An important example is the meridional circulation, which is also
seen on the solar surface in Doppler-velocity observations
\citep[e.g.,][]{Hathaw1996}, such that the flow is generally in the
pole-ward direction.
Using time-distance analysis \cite{Giles1997} found a similar flow
in the upper parts of the convection zone.
This was analysed in more detail by \cite{Haber2002} and \cite{Zhao2004}
using ring-diagram and time-distance analyses, respectively.
On smaller scales local analyses have revealed convective flows 
on supergranular scales, as well as complex flows associated
with active regions (see also Kosovichev, these proceedings).

\section{Open questions}


While major progress has obviously been made on our knowledge about
rotation in the solar interior, much remains to be done.
The determination of the rotation of the inner solar core is still highly
uncertain.
Unfortunately, substantial improvement in the low-degree p-mode splittings,
required to determine the core rotation, will require much more extensive
observations than now available;
observation of g modes, which are more sensitive to the properties
of the solar core, remains elusive despite recent encouraging 
progress \citep{Garcia2006}.
Better observational constraints are also required on the properties of
the tachocline region; this includes details of solar structure 
such as the depth of the convective envelope and the detailed properties
of the transition to the radiative region, as well as a better determination
of the location and thickness of the transition in rotation.
Also, it is obviously important to test the reality of the 1.3~yr oscillation
in rotation in the tachocline region and 
investigate its diagnostic potential.

From a theoretical point of view, the evolution of rotation in the
radiative interior of the Sun is still poorly understood, although
there are models that are able to reproduce the present near-constant
rotation through, for example, the effect of a dynamo-generated
magnetic field \citep{Eggenb2005} or transport by gravity waves
\citep{Charbo2005}.
This issue is obviously closely related to the establishment of
the tachocline, where magnetic fields have also been invoked
as the most likely mechanism \citep[e.g.,][]{Gough1998}.
Progress towards understanding
the mean rotation profile within the convection zone has been made in terms
of mean-field models which involve thermal effects in the tachocline
region \citep{Rempel2005}.
Evidently, the preferred route to understand the dynamics of the convection
zone would be through realistic three-dimensional simulations.
\cite{Brun2002} showed that increased numerical resolution in such
simulations improved the agreement with the observed profile,
and the simulations
have confirmed that including entropy variations in the tachocline region
can act to bring the results into better agreement with the helioseismic
inferences \citep[][see also Brun {\etal}\ these proceedings]{Miesch2006}.
Interestingly, detailed modelling of the outer 5 \% of the solar radius
\citep{DeRosa2002}
has reproduced a near-surface decrease in the angular velocity,
such as observed in the shear layer (cf.\ Figure~\ref{fig:surshear}).
A definite model of the zonal flows is still lacking; one may hope
that such flows will eventually emerge as natural features
of the hydrodynamical simulations.

\begin{acknowledgments}
I am very grateful to T. Corbard and R. Howe for help with figures in this
review.
\end{acknowledgments}

\begin{discussion}

\discuss{Pecker}{
The trend towards the equator {\em seems} to be much quicker (in
your diagram) for torsional oscillations than it is for helioseismological
data. Why is that so?
}

\discuss{Christensen-Dalsgaard}{
In fact, my impression from Figure~\ref{fig:zonallat} is that the
helioseismically inferred flow is quite similar to the behaviour of
the surface torsional oscillation and, if anything, is at slightly lower
latitude.
This could be related to the apparent propagation towards the surface,
as illustrated in Figure~\ref{fig:zonalrad}.
}

\end{discussion}

\end{document}